\documentclass[twocolumn]{article}
\usepackage{times,bm}
\usepackage{amsmath}
\usepackage[affil-it,]{authblk}
\usepackage[pagebackref=false,colorlinks=true,linkcolor=acsblue,citecolor=olive,urlcolor=acsblue]{hyperref}

\usepackage{geometry}

\usepackage[switch*, displaymath,pagewise, mathlines]{lineno}
\leftlinenumbers
\usepackage{graphicx}
\usepackage{cite}
\usepackage{amsmath}
\usepackage{amsfonts}
\usepackage{amssymb}
\usepackage{slashed}
\usepackage{subfloat}
\usepackage{slashed}
\usepackage{color}
\usepackage{float}
\usepackage{multirow,multicol}
\usepackage{tabulary}
\usepackage[flushleft]{threeparttable}
\usepackage{pgfplots,subfigure}
\usepackage{xcolor}
\numberwithin{equation}{section}
\usepackage{tikz}
\usepackage{ulem}
\usepackage{hyperref}
\usepackage{nameref}
\usepackage{comment}

\usepgflibrary{arrows}
\usetikzlibrary{shapes.callouts}
\tikzset{
	level/.style   = { thick, },
	connect/.style = { dotted, red   },
	notice/.style  = { draw, rectangle callout, callout relative pointer={#1} },
	label/.style   = { text width=1cm }
}

\definecolor{acsblue}{RGB}{17,76,139}
\definecolor{shadecolor}{RGB}{255,241,204}
\usepackage{letltxmacro}
\makeatletter
\let\oldr@@t\r@@t
\def\r@@t#1#2{%
	\setbox0=\hbox{$\oldr@@t#1{#2\,}$}\dimen0=\ht0
	\advance\dimen0-0.2\ht0
	\setbox2=\hbox{\vrule height\ht0 depth -\dimen0}%
	{\box0\lower0.4pt\box2}}
\LetLtxMacro{\oldsqrt}{\sqrt}
\renewcommand*{\sqrt}[2][\ ]{\oldsqrt[#1]{#2}}
\makeatother
\usepackage{environ}

\begin{document}

\newcommand{{\ri}}{{\rm{i}}}
\newcommand{{\Psibar}}{{\bar{\Psi}}}
\newcommand*\var{\mathit}
%
%
\fontsize{8}{10}\selectfont

\title{\mdseries{Damped photonic modes in helical graphene}}

\author{ \textit {\mdseries{Abdullah Guvendi}}$^{\ 1}$\footnote{\textit{ E-mail: abdullah.guvendi@erzurum.edu.tr  } }~,~ \textit {\mdseries{Omar Mustafa}}$^{\ 2}$\footnote{\textit{ E-mail: omar.mustafa@emu.edu.tr  } }~,~ \textit {\mdseries{Abdulkerim Karabulut}}$^{\ 1}$\footnote{\textit{ E-mail: akerimkara@gmail.com (Corr. Auth.) } } \\
	\small \textit {$^{\ 1}$ \footnotesize Department of Basic Sciences, Erzurum Technical University, 25050, Erzurum, Türkiye}\\
	\small \textit {$^{\ 2}$ \footnotesize Department of Physics, Eastern Mediterranean University, 99628, G. Magusa, north Cyprus, Mersin 10 - Türkiye}}
 
\date{}
\maketitle

\begin{abstract}
We analyze the behavior of spin-1 vector bosons in helical spacetime, focusing on photonic modes in helical graphene structures. We model the helical graphene surface as a smooth, continuous, and distortion-free manifold, effectively adopting the continuum approximation. By solving the fully covariant vector boson equation, we derive exact solutions that describe the quantum states of photons in a curved helical background, revealing their energy spectra, mode profiles, and decay dynamics. We find that the decay times of damped photonic modes range from \(10^{-16}\) to \(10^{-13}\) seconds as the helical pitch (\(a\)) varies from \(10^3\) nanometers to \(1\) nanometer, indicating that the structure efficiently absorbs all photonic modes. Additionally, the probability density functions exhibit time dependence, complementing their spatial variation. These findings provide a foundation for the design of ultrafast graphene photodetectors, graphene photodevices for high-speed optical communications, advanced photonic devices, and quantum materials based on helical graphene for various nanophotonic applications.
\end{abstract}

\begin{small}
\begin{center}
\end{center}
\end{small}

\section{\mdseries{Introduction}}\label{sec1}

The study of quantum fields in curved spacetime has gained prominence, particularly for spin-1 particles, which are governed by the Duffin-Kemmer-Petiau (DKP) equation-a first-order relativistic wave equation describing both spin-0 and spin-1 particles \cite{1,2,3}. The DKP equation has found extensive applications across various fields, including quantum chromodynamics \cite{4}, covariant Hamiltonian dynamics \cite{5,6}, and scattering phenomena \cite{7}. Investigations of the DKP equation in curved spacetimes and Riemann-Cartan geometries have revealed novel aspects of spin-1 particle dynamics, particularly in the massless particle limit. In this regime, the DKP equation reduces to the Maxwell equations in lower dimensions, providing valuable insights into the behavior of massless vector fields. A key motivation for studying the DKP equation lies in its ability to describe spin-1 particles, such as vector bosons, in curved spacetime. In \(2+1\)-dimensions, the vector boson equation, representing the spin-1 sector of the DKP framework, has been derived through the canonical quantization of Barut’s classical zitterbewegung model \cite{10,11,12,nuri}. This equation is especially useful for examining relativistic dynamics in lower dimensional non-trivial spacetimes \cite{13,14,15,16}, where spacetime curvature significantly impacts particle behavior, altering their propagation characteristics and interaction dynamics \cite{book1}.

\vspace{0.15cm}
\setlength{\parindent}{0pt}

On the other hand, helical spacetime, characterized by a spiraling surface, offers an interesting background for investigating the dynamics of spin-1 particles. A helicoid, with its spiral and radial parameters, provides a mathematically rich and physically relevant model for systems like helical graphene \cite{17,18,19}. This geometry significantly modifies the quantum dynamics of particles, and the solutions of the vector boson equation in such a background offer insight into how curvature influences energy spectra, decay dynamics, and mode profiles. Graphene, a significant member of the two-dimensional material family, consists of a single layer of carbon atoms arranged in a hexagonal lattice, only one atom thick. This exceptional atomic arrangement imparts graphene with outstanding chemical, physical, and electrical properties \cite{tarhini2023advances, kumar2023electronic}. Despite its thinness, graphene exhibits an unusual photon absorption characteristic, as thin materials typically have low optical absorption. Graphene can absorb light across a broad spectral range, from visible light to terahertz frequencies. Single-layer graphene, which is approximately 97.7 $\%$ transparent, absorbs around 2.3 $\%$ of visible light-considered relatively high for optical devices. Its broad-spectrum absorption enables applications in infrared detection, solar cells, and wideband photodetectors \cite{nair2008fine}. When multiple layers of graphene are stacked, the absorption increases linearly. The photonic absorption of graphene stems directly from its unique electronic band structure. In graphene, electrons behave like massless particles within the “Dirac cones,” a unique energy band structure that facilitates the absorption of photons across a wide energy spectrum, making graphene responsive to both low- and high-energy photons \cite{mak2008measurement}. The sp² hybridization of bonds between carbon atoms creates a structure that allows electrons to move with minimal resistance, giving single-layer graphene its remarkably high electrical conductivity \cite{novoselov2004electric}. The rapid, nearly resistance-free transport of electrons through the graphene layer minimizes energy loss, making graphene ideal for photodetectors that require swift response times \cite{xia2009ultrafast}. Additionally, although extremely thin, graphene is remarkably strong and flexible, exhibiting approximately 200 times the strength of steel \cite{lee2008measurement}. Its structural flexibility and transparency make it suitable for transparent and flexible electrodes, crucial for wearable devices, flexible sensors, and bendable displays \cite{wang2008transparent}. Twisted monolayer graphene exhibits modifiable optical properties, such as enhanced absorption and tunable plasmonic resonances, which are advantageous for applications in optoelectronics. The twist angle exerts a direct influence on these properties, facilitating in-depth exploration of band structure and carrier dynamics, and positioning twisted graphene as a promising candidate for use in photodetectors, tunable lasers, and similar advanced devices \cite{do2022generalized}. Besides, single-layer graphene boasts a surface area of 2630 m$^{2}$/g, allowing it to serve effectively in energy storage devices like batteries and supercapacitors \cite{stoller2008graphene}. Graphene gains extraordinary electronic and optical properties, exhibits novel phenomena when shaped into helical structures \cite{20,21}. Curvature effects in such configurations modify the electronic band structure, enabling the emergence of topologically protected states, tunable bandgaps, and enhanced light-matter interactions. These properties make helical graphene highly valuable for various applications, including quantum computing, optoelectronics, and nanotechnology \cite{21}. The ability to confine and manipulate photonic modes through helical graphene’s unique geometry is particularly promising for the development of advanced photonic devices \cite{22}. The manipulation of photonic modes in helical graphene holds significant practical implications for optical communications, quantum information processing, and the design of photonic circuitry. Helical graphene can enable precise control of photonic states, making it ideal for designing new devices \cite{23,24,25,26}. This ability to engineer photonic modes through geometric parameters has the potential to revolutionize the design of metasurfaces and nanophotonic devices, making it a cornerstone of future advancements in nanophotonics and quantum technologies. Although numerous experimental and theoretical studies have explored the properties of graphene, no analytical model describing the photonic modes of helical graphene has been reported in the literature. This study aims to fill that gap. We model the helical surface as a smooth, continuous, and distortion-free manifold by adopting a continuum approach. This allows us to isolate and analyze the curvature-induced effects on photonic modes, intentionally neglecting atomic-scale characteristics such as lattice discreteness and graphene’s intrinsic periodic potential. Within this framework, we investigate the relativistic quantum dynamics of spin-1 particles in a helical spacetime, focusing on the time evolution of massless spin-1 vector bosons (photons) confined to a helical graphene structure. Our objective is to obtain analytically exact solutions to the corresponding fully covariant vector boson equation.

\vspace{0.15cm}
\setlength{\parindent}{0pt}

This manuscript is organized as follows: In section \ref{sec2}, we introduce the concept of helical spacetime and the associated vector boson equation. Section \ref{sec3} discusses the resulting second-order wave equation and explores analytically allowed solutions under the assumption of a narrow helicoid. The analytical solutions, formulated in terms of confluent hypergeometric functions, enable us to determine the wave functions and energy profiles. We find that the quantum states for photons decay over time without exhibiting real oscillation, which facilitates the calculation of decay times. Consequently, we derive the time-dependent wave functions and probability density functions. Finally, section \ref{conc} provides a summary and conclusions of our findings.

\section{\mdseries{Vector bosons in helical spacetime}} \label{sec2}

\begin{figure}[h!]
\centering
\includegraphics[scale=0.50]{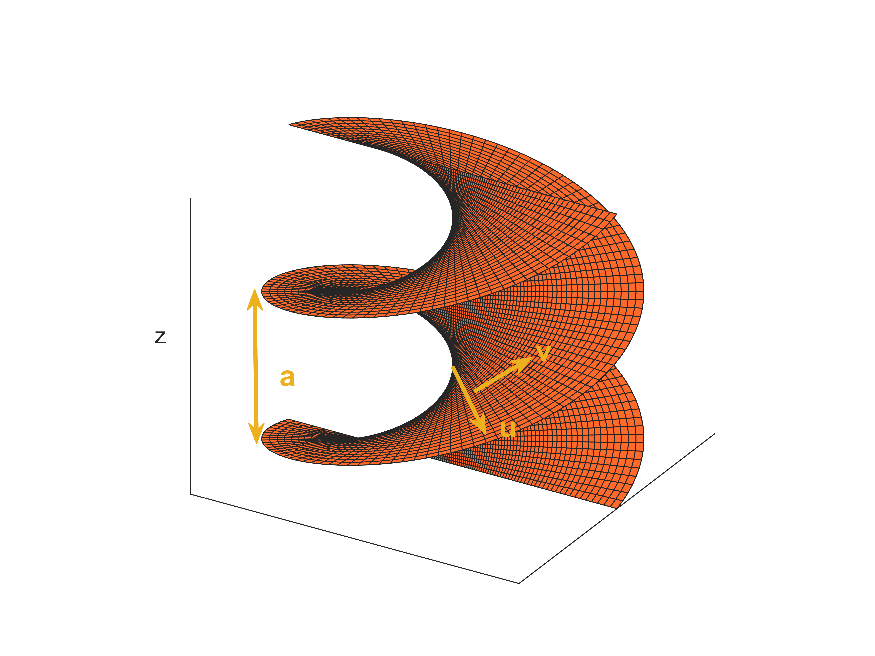}
\caption{\footnotesize Visualization of a helicoid aligned with the z-axis in three-dimensional Euclidean space. The surface displays the isothermal coordinates \( (u, v) \), with arrows indicating their respective directions.}
\label{fig:1}
\end{figure}

In this section, we present the helical surface and derive the necessary operators for the corresponding vector boson equation. We begin by introducing the helical spacetime, which can be characterized through the parametrization outlined in \cite{17}
\begin{eqnarray}
\vec{r}\left(u,v\right)=a\sinh(u)\cos(v)\hat{i}+a\sinh(u)\sin(v)\hat{j}+av\hat{k},
\end{eqnarray}
where the coordinate $v$ spirals about the axis of the helicoid, $u$ shoots out normal to $v$ from the axis with $a$ the pitch of the spiral. A helicoid in three-dimensional Euclidean
space can be seen in the Figure \ref{fig:1}.
Given this parametrization, one finds
\begin{equation*}
\begin{split}
dx &= \frac{\partial x}{\partial u}du+\frac{\partial x}{\partial v}dv\\
&\Rightarrow a\cosh(u)\cos(v)du-a\sinh(u)\sin(v)dv,\\
dy &= \frac{\partial y}{\partial u}du+\frac{\partial y}{\partial v}dv\\
&\Rightarrow a\cosh(u)\sin(v)du+a\sinh(u)\cos(v)dv,\\
dz &= adv.
\end{split}
\end{equation*}
Accordingly, spatial part of the line element ($d\tilde{s}^{2}$) can be determined as the following;
\begin{equation*}
d\tilde{s}^{2}=dx^2 + dy^2 + dz^2 \Rightarrow a^2\cosh^2(u)\left[du^2+dv^2\right].\label{eq1}
\end{equation*}
By trivially projecting the temporal coordinate (\( t \)) from flat space-time, where the helicoid resides, we can describe the helical background spacetime through the following \((2+1)\)-dimensional curved spacetime metric ($ds^2$) with the signature $(+,-,-)$ 
\begin{eqnarray}
ds^2=c^2dt^2-\xi^2\left(u\right)\left[du^2+dv^2\right], \label{eq2}
\end{eqnarray}
where \( \xi\left(u\right)=a\cosh(u) \), and \( c \) is the speed of light. Based on the line element in Eq. (\ref{eq1}), the covariant metric tensor ($g_{\mu\nu}$) and its inverse ($g^{\mu\nu}$) can be expressed as
\begin{equation*}
\begin{split}
&g_{\mu \nu}=\text{diag}\left(c^2, -\xi^2\left(u\right), -\xi^2\left(u\right)\right),\quad g^{\mu \nu}=\text{diag}\left(\frac{1}{c^2}, -\frac{1}{\xi^{2}\left(u\right)}, -\frac{1}{\xi^{2}\left(u\right)}\right),\\
&g^{\mu \nu}g_{\mu \nu}=\textbf{I}_3,\quad (\mu,\nu=t,u,v), \label{eq3}
\end{split}
\end{equation*}
where $\textbf{I}_3$ denotes three dimensional identity matrix. Now, let us introduce the fully-covariant vector boson equation. In $(2+1)$-dimensional curved spacetime, this equation can be expressed as the following \cite{13}:
\begin{equation}
\left(\mathcal{B}^{\mu} \slashed{\nabla}_{\mu} + i \tilde{m} \textbf{I}_4\right)\Psi(x^{\mu})=0,\label{eq4}
\end{equation}
where $\tilde{m} = \frac{mc}{\hbar}$, with $m$ representing the rest mass of the vector boson, $\hbar$ is the usual (reduced) Planck constant, $\slashed{\nabla}_{\mu}$ denotes the covariant derivative, with $\slashed{\nabla}_{\mu} = \partial_{\mu} - \Omega_{\mu}$. Here, $\Psi$ represents the symmetric rank-two spinor constructed as the direct (Kronecker) product of two symmetric Dirac spinors. $\mathcal{B}^{\mu}(x^{\mu})$ matrices are spacetime-dependent spin-1 matrices derived from the generalized Dirac matrices $\gamma^{\mu}(x^{\mu})$ and are expressed as the following \cite{13}:
\begin{equation*}
\mathcal{B}_{\mu}(x^{\mu}) = \frac{1}{2} (\gamma^{\mu}(x^{\mu}) \otimes \textbf{I}_{2} + \textbf{I}_{2} \otimes \gamma^{\mu}(x^{\mu})),
\end{equation*}
The vector $x^{\mu}$ denotes the spacetime position vector. Also, $\textbf{I}_d$ indicates $d$-dimensional identity matrix. The spacetime-dependent Dirac matrices can be determined by means of the free Dirac matrices ($\gamma^{k}$, $k=0,1,2.$) through the relation $\gamma^{\mu}=e^{\mu}_{k}\gamma^{k}$ where $e^{\mu}_{k}$ are inverse tetrad fields. The space-independent (free) Dirac matrices are chosen by using the Pauli spin matrices (\(\sigma_{x}, \sigma_{y}, \sigma_{z}\)) as the following: \( \gamma^{0}=\sigma_{z} \), \( \gamma^{1}=i\sigma_{x} \), \( \gamma^{2}=i\sigma_{y} \), according to the signature (\(+,-,-\)) of the line element. Here $i=\sqrt{-1}$ and $\sigma^2_{x(yz)}=\textbf{I}_2$. In Eq. (\ref{eq4}), $\Omega_\mu$ are the spinorial affine connections for the spin-1 field, and are determined from the affine spin connections ($\Gamma_\mu$) for the Dirac fields as \cite{13}:
\begin{equation*}
\Omega_{\mu} = \Gamma_{\mu} \otimes \textbf{I}_{2} + \textbf{I}_{2} \otimes \Gamma_{\mu}.
\end{equation*}
The spinorial affine connections for the Dirac field, denoted \( \Gamma_{\lambda} \), are given by: \(\Gamma_{\lambda} = \frac{1}{4} g_{\mu \tau} \left[e^{k}_{\nu,\lambda} e^{\tau}_{k} - \Gamma_{\nu \lambda}^{\tau} \right] \mathcal{S}^{\mu \nu}  \), where $,\lambda$ denotes differentiation with respect to \( x^{\lambda} \) and tetrad fields are denoted by \( e^{k}_{\tau} \). Here, \( \Gamma_{\nu \lambda}^{\tau} \) represents the Christoffel symbols, defined as: \( \Gamma_{\nu \lambda}^{\tau} = \frac{1}{2} g^{\tau \epsilon} \left[\partial_{\nu} g_{\lambda \epsilon} + \partial_{\lambda} g_{\epsilon \nu} - \partial_{\epsilon} g_{\nu \lambda} \right] \) \cite{27,28}. The \( \mathcal{S}^{\mu \nu} \) represents the spin operators, defined by: \( \mathcal{S}^{\mu \nu} = \frac{1}{2} \left[\gamma^{\mu}, \gamma^{\nu} \right] \) \cite{27,28}. Greek indices refer to the coordinates in the curved spacetime, while Latin indices refer to the coordinates in flat Minkowski spacetime. The tetrads and their inverses are determined using the equations \( g_{\mu\nu} = e^{k}_{\mu} e^{l}_{\nu} \eta_{kl} \) and \( e^{\mu}_{k} = g^{\mu\nu} e^{l}_{\nu} \eta_{kl} \), where \( \eta_{kl} \) is the flat Minkowski tensor given by \( \eta_{kl} = \textrm{diag}(1, -1, -1) \) \cite{27,28}. From this, we obtain the following results:
\begin{flalign*}
&e^{k}_{\mu} = \textrm{diag}(c, a\cosh(u), a\cosh(u)),\\
& e^{\mu}_{k} = \textrm{diag}\left(\frac{1}{c}, \frac{1}{a\cosh(u)}, \frac{1}{a\cosh(u)}\right).
\end{flalign*}
Consequently, we find \( \gamma^{t} = \frac{1}{c} \sigma_{z} \), \( \gamma^{u} = \frac{i}{a\cosh(u)} \sigma_{x} \), and \( \gamma^{v} = \frac{i}{a\cosh(u)} \sigma_{y} \). The non-zero components of the Christoffel symbols are calculated as follows: \( \Gamma_{uu}^{u} = \tanh(u) \), \( \Gamma_{vv}^{u} = -\tanh(u) \), and \( \Gamma_{uv}^{v} = \tanh(u) \). Additionally, the non-zero component of the spinorial affine connection is given by \( \Gamma_{v} = \frac{i}{2} \tanh(u) \sigma_{z} \).

\section{\mdseries{Wave equation and its solutions}} \label{sec3}

\begin{figure}[ht]
\centering
\includegraphics[scale=0.50]{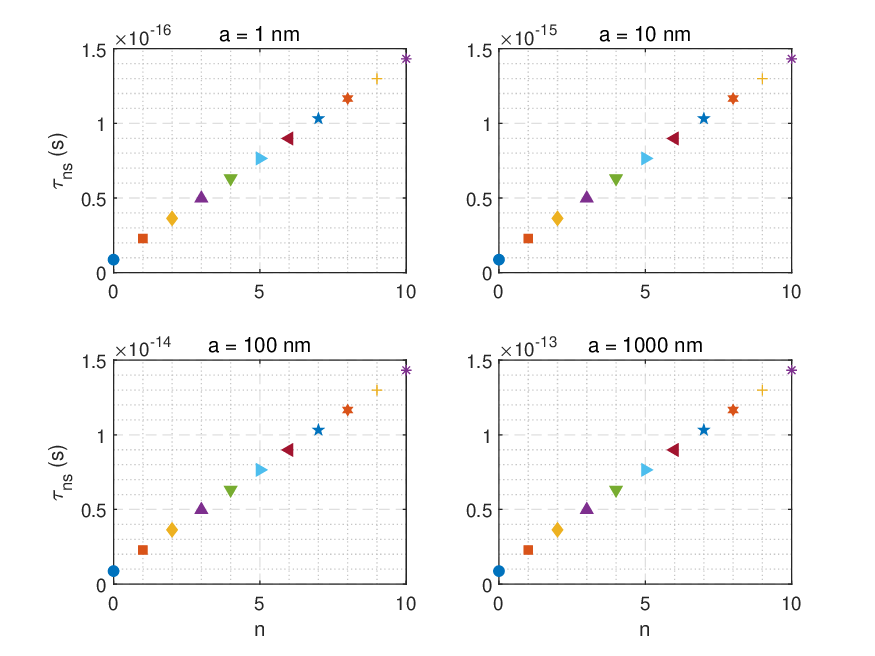}
\caption{\footnotesize Decay time (\(\tau_{ns}\)) of damped photonic modes for various pitch (\(a\)) values. Here $s=1$.}
\label{fig:2}
\end{figure}

In this section, we derive coupled equations for relativistic spin-1 bosons in helical spacetime and explore analytically feasible solutions for the resulting wave equation. Based on the line element in Eq. (\ref{eq2}), we decompose the spin-1 field as follows:  
\begin{equation*}
\Psi(x^{\mu})=\textrm{e}^{-i\frac{E}{\hbar}\,t}\textrm{e}^{i\,s\,v}\left(\psi_{1}(u),\psi_{2}(u),\psi_{3}(u),\psi_{4}(u)\right)^{T},
\end{equation*}
where \(E\) is the relativistic energy, \(s\) is the spin, and \(^{T}\) denotes the transpose of the \(u\)-dependent spinor. Following some algebraic manipulation detailed in \cite{13,14}, we arrive at the following results:
\begin{equation}
\begin{split}
&\tilde{\mathcal{E}}\psi_{+}(u)-\tilde{m}\psi_{-}(u)-\frac{s}{\xi(u)}\psi(u)=0,\\
&\tilde{\mathcal{E}}\psi_{-}(u)-\tilde{m}\psi_{+}(u)-\frac{\dot{\psi}(u)}{\xi(u)}=0,\\
&\tilde{m}\psi(u)+\frac{\dot{\psi_{+}}(u)}{\xi(u)}+\frac{\dot{\xi}(u)}{\xi(u)^2}\psi_{+}(u)-\frac{s}{\xi(u)}\psi_{-}(u)=0,\label{Eqq}
\end{split}
\end{equation}
where \(\tilde{\mathcal{E}}=\frac{E}{\hbar c}\), the dot denotes the derivative with respect to \(u\), \(\psi_{+}(u)=\psi_{1}(u)+\psi_{4}(u)\), \(\psi_{-}(u)=\psi_{1}(u)-\psi_{4}(u)\), and \(\psi(u)=\psi_{2}(u)+\psi_{3}(u)\) since \(\psi_{2}(u)=\psi_{3}(u)\). Solving this system for \(\psi(u)\), we obtain the following wave equation:
\begin{equation}
\ddot{\psi}(u)+\left[a^2\cosh^2(u)\left(\tilde{\mathcal{E}}^2-\tilde{m}^2\right)-s^2\right]\psi(u)=0,\label{om1}
\end{equation}
which describes massive vector bosons. A complete solution to this wave equation is challenging (see also \cite{17}); however, considering a narrow helicoid, we approximate \(\cosh(u)^2 \approx 1 + u^2 + \mathcal{O}(u^4)\). This leads to:
\begin{eqnarray}
\ddot{\psi}(u)+\left[\tilde{\lambda} - \Omega^2 u^2\right]\psi(u)=0,
\end{eqnarray}
where \(\Omega^2 = a^2\left(\tilde{m}^2 - \tilde{\mathcal{E}}^2\right) = a^2\left(\tilde{m}^2 + \eta^2\right)\), \(\eta = i\tilde{\mathcal{E}} \in \Re\), and \(\tilde{\lambda} = -(s^2 + \Omega^2)\). This equation has an exact textbook solution in the form of a confluent hypergeometric function/series \cite{27,stegun}:
\begin{equation}
    \psi(u)=\mathcal{N}\, u\, \exp\left(-\frac{\Omega u^2}{2}\right) \, _1F_{1} \left(\frac{3}{4} - \frac{\tilde{\lambda}}{4\Omega}, \frac{3}{2}, \Omega u^2\right).
\end{equation}
The confluent hypergeometric series must be truncated to a polynomial of order \(n \geq 0\) \cite{27} by the condition \(\frac{3}{4} - \frac{\tilde{\lambda}}{4\Omega} = -n \Rightarrow \tilde{\lambda} = 4\Omega \tilde{n}\,;\,\, \tilde{n} = n + \frac{3}{4}\). This implies \(\Omega = -2\tilde{n} + \sqrt{4\tilde{n}^2 - s^2}\). For photons (\(\tilde{m} = 0\)) \cite{13}, this yields:
\[
\Omega = -ia\tilde{\mathcal{E}} = -a\eta = -2\tilde{n} + \sqrt{4\tilde{n}^2 - s^2},
\]
so that
\[
a\eta_{ns} = 2\tilde{n} - \sqrt{4\tilde{n}^2 - s^2} > 0
\]
where \( \eta_{ns} \in \Re\), which leads to:
\begin{equation}
E_{ns} = -\frac{i\hbar c}{a}\left[2\tilde{n} - \sqrt{4\tilde{n}^2 - s^2}\right].
\end{equation}
Thus, the corresponding wave function is:
\begin{equation}
    \psi_{ns}(u)=\mathcal{N}\, u\, \exp\left(-\frac{a\eta_{ns} u^2}{2}\right) \, _1F_{1} \left(-n, \frac{3}{2}, a\eta_{ns} u^2\right).
\end{equation}
Note that \(a\eta_{ns} > 0\) for all \(\tilde{n} \geq \frac{3}{4}\), ensuring the asymptotic convergence \(\psi_{ns}(u) \rightarrow 0\) as \(u = 0\) and \(u = \infty\), making the wave function finite and square-integrable. For the photon (\(s = 1\)) ground state (\(n = 0\)), we find \cite{stegun}:
\begin{equation}
    \psi_{01}(u) = \mathcal{N}\, u\, \exp\left(-\frac{a\eta_{01} u^2}{2}\right)\,; \,\, \eta_{01} = \frac{1}{2a}(3 - \sqrt{5}),
\end{equation}
and:
\begin{equation}
    \psi_{01}(t,u,v) = \mathcal{N} u\, \exp\left(-\frac{a\eta_{01} u^2}{2}\right)e^{iv}\exp\left(-\frac{c}{a}\eta_{01}\,t\right).
\end{equation}
The time-dependent probability density for the photonic ground state is:
\begin{equation}
\begin{split}
&P_{01}(t,u) =\\
&|\mathcal{N}|^2\,u^2 \exp\left(-\frac{1}{2}(3 - \sqrt{5}) u^2\right) \exp\left(-\frac{c}{a^2}(3 - \sqrt{5})\,t\right).
\end{split}
\end{equation}
Here, note that \(a\) has units of length. This result implies that the corresponding states are not steady and decay over time with decay time \(\tau_{ns} = \frac{\hbar}{|\Im E_{ns}|}\), where \(\Im E_{ns}\) is the imaginary part (negative) of the energy \cite{13}:
\begin{equation}
\tau_{ns} = \frac{a}{c\left[2\tilde{n} - \sqrt{4\tilde{n}^2 - s^2}\right]},\label{DT}
\end{equation}
since \(\Psi \propto \exp(-i\frac{E_{ns}}{\hbar}t)\). The decay time for various values of \(a\) is shown in Figure \ref{fig:2}. These results suggest that the decay times of damped photonic modes is influenced by the helical surface structure ($\propto a$). Specifically, larger pitch values increase spacing between turns, reducing interactions and extending decay times, while smaller pitch values enhance confinement, resulting in shorter decay times due to faster energy dissipation. The pitch thus plays a crucial role in determining the balance between mode confinement and energy dissipation, affecting photonic mode stability. Additionally, higher-energy modes exhibit relatively longer decay times, while lower-energy modes decay more quickly. 

\begin{figure}[ht]
\centering
\includegraphics[scale=0.50]{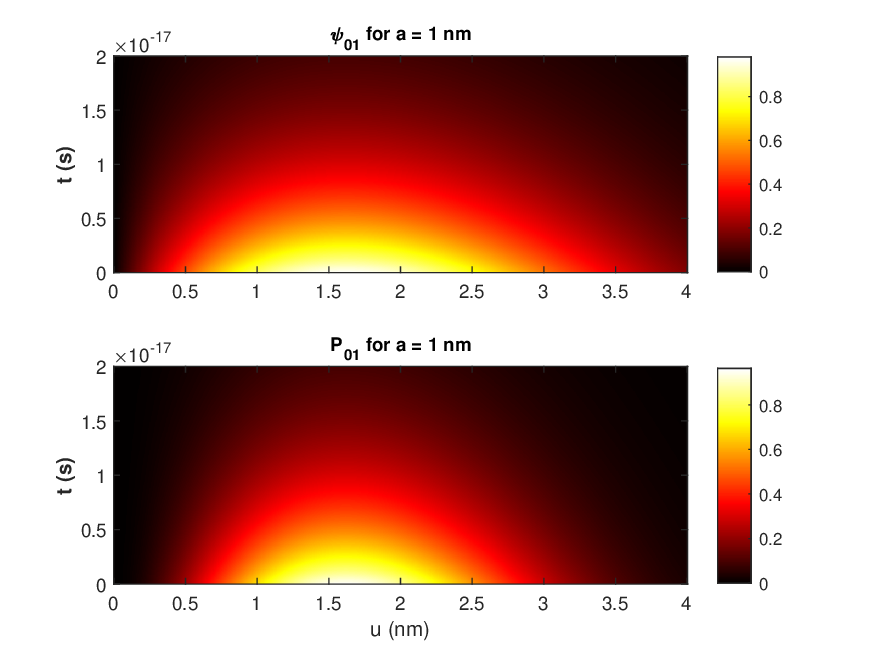}
\caption{\footnotesize This figure presents the wave function and the probability density function as functions of time \(t\) (vertical axis) and position \(u\) (horizontal axis). The calculations were performed using \(c=3\times10^{17} \) (nm/s) . The color intensity represents the magnitude of the wave function and probability density, illustrating the temporal evolution and spatial distribution of the quantum state.}
\label{fig:3}
\end{figure}

Figure \ref{fig:3} shows the ground state wave function and the corresponding probability density function as functions of time and position. The decrease in the probability density function over time indicates that the photonic modes in the helical graphene structure are undergoing a form of decay. This could imply that the energy or intensity of these modes diminishes as time progresses, due to dissipative processes or interactions with the environment, leading to a reduction in the likelihood of photon localization within the system. 

\section{\mdseries{Summary and discussions}}\label{conc}

This paper explores the dynamics of spin-1 vector bosons, specifically focusing on photonic modes in helical graphene structures, by solving the fully covariant vector boson equation. We derive coupled equations that describe the quantum states of these particles within the curved geometry of a helical spacetime. Analytical solutions are presented for narrow helicoid geometries, expressed through confluent hypergeometric functions, shedding light on the key characteristics of the energy spectra, mode profiles, and decay dynamics of the system.

\vspace{0.15cm}
\setlength{\parindent}{0pt}

A significant finding is the impact of the helical surface geometry, particularly the pitch and radial extension, on the behavior of photonic modes. Larger pitch values, corresponding to broader helical turns, result in longer decay times due to weaker confinement, allowing photonic modes to dissipate energy more gradually. Conversely, smaller pitch values lead to stronger confinement and more intense interactions within the helicoid, accelerating energy dissipation and shortening decay times. This relationship between pitch and decay time provides critical insight into how geometric parameters influence the behavior of photonic modes in such structures. Moreover, the energy spectra indicate that higher-energy modes exhibit relatively longer decay times, while lower-energy modes decay more rapidly, emphasizing the interplay between energy levels and confinement. We observe that the decay times of damped photonic modes range from \(10^{-16}\) to \(10^{-13}\) seconds as the helical pitch (\(a\)) varies from \(10^3\) nanometers to \(1\) nanometer, highlighting the structure's ability to effectively absorb all photonic modes (see also \cite{new}). The absorption time for ultrafast graphene photodetectors is typically in the femtosecond range (see also \cite{new2}), primarily due to the high electron mobility and ultrafast dynamics of photoexcited carriers in graphene. When light is absorbed by graphene, electron-hole pairs are generated almost instantaneously. The carrier relaxation time, which refers to the period during which the carriers lose excess energy, ranges from 100 fs to 1 ps \cite{guvendi-epjp}. Although the intrinsic optical absorption of graphene is relatively low, it can be enhanced through geometric deformations, improving the overall efficiency. The full response time, which includes carrier transport to the electrodes, can extend to the picosecond range, but the absorption itself remains very fast, typically occurring in the fs timescale.

\vspace{0.15cm}
\setlength{\parindent}{0pt}

Our findings significantly advance the theoretical understanding of quantum fields in curved spacetimes and provide valuable insights for the design of photonic devices and quantum materials. By precisely tuning the geometric parameters of the helicoid, it becomes possible to manipulate photonic mode confinement and dissipation, thereby unlocking new opportunities in nanophotonics and contributing to the development of next-generation quantum materials and photonic technologies based on helical graphene architectures.

\vspace{0.15cm}
\setlength{\parindent}{0pt}

In this study, we model the helical graphene surface as a smooth, continuous, and distortion-free manifold, adopting a continuum approach. This enables a focused analysis of curvature-induced effects on photonic modes while deliberately excluding atomic-scale features such as lattice discreteness and the periodic potential intrinsic to graphene. Our central objective is to elucidate the impact of global geometric deformations on photon dynamics through the application of a fully covariant vector boson equation. The helical graphene configuration considered here represents a distinctive class of low-dimensional materials, wherein a graphene strip undergoes continuous helical deformation, forming a screw-like geometry. Within this framework, we derive analytical solutions that clearly demonstrate how curvature governs photonic behavior. A rigorous treatment of lattice-level effects, including electronic band structure and hopping variations, is beyond the scope of the present work and is reserved for future investigations.

\section*{\small{Acknowledgment}}

The authors sincerely thank the Editor and the reviewer for their valuable comments, insightful suggestions, and thoughtful questions, which have significantly enhanced the clarity, readability, and overall quality of the manuscript.

\section*{\small Credit authorship contribution statement}

\textbf{Abdullah Guvendi}: Conceptualization, Methodology, Formal Analysis, Writing – Original Draft, Investigation, Visualization, Writing – Review and Editing.\\
\textbf{Omar Mustafa}: Conceptualization, Methodology, Formal Analysis,  Writing – Original Draft, 
Investigation, Writing – Review and Editing, Project Administration.\\
\textbf{Abdulkerim Karabulut}: Conceptualization, Methodology, Formal Analysis, Validation, Investigation, Writing – Original Draft, Writing – Review and Editing.

\section*{\small{Data availability}}
This is a theoretical research, including equations and derived quantities presented in the main text of the paper.

\section*{\small{Conflicts of interest statement}}
The authors have disclosed no conflicts of interest.

\section*{\small{Funding}}
This research has not received any funding.

\end{document}